\documentclass{article}
\usepackage{amsmath}
\usepackage{amsfonts}
\usepackage{amssymb}
\usepackage{amsthm}
\theoremstyle{definition}
\newtheorem{theorem}{Theorem}
\newtheorem{definition}{Definition}

\begin{document}
\title{Classical Fermionic Dynamics}
\author{Luther Rinehart\\ University of Pittsburgh\\ ldr22@pitt.edu}
\maketitle

\begin{abstract}
The formulation of classical mechanics applicable to fermionic degrees of freedom is presented in mathematically rigorous terms, including a description of how the mathematical structure relates to the quantization of the theory. Canonical transformations and the algebra of observables are defined and studied. A formula is given for the analog of the Poisson bracket. The quantization of the theory proceeds according to deformation quantization.
\end{abstract}

\section{Intoduction}
In the past century, it was discovered that classical mechanics, via its Hamiltonian formulation, is described by symplectic geometry\cite{Symplectic}. Phase space is a symplectic vector space (or manifold), meaning that it has defined on it a  particular antisymmetric non-degenerate bilinear form. Dynamical evolution is simply a one-parameter family of diffeomorphisms of this space such that preserve this symplectic form. It can be shown that such a dynamical evolution is generated by some smooth real-valued function $H$ on phase space, called the Hamiltonian.  More generally, there is an algebra of observables consisting of the smooth real-valued functions on phase space.\\
Elements of such a symplectic vector space are called \emph{bosonic} degrees of freedom.  The classical mechanics of bosonic degrees of freedom as well as their quantum mechanics are widely studied and well understood.  It was soon discovered that there exist other types of degrees of freedom in nature called \emph{fermionic}, which do not live in a symplectic vector space, but rather in an orthogonal vector space, one with a \emph{symmetric} non-degenerate bilinear form. The quantum mechanics of fermions have become a crucial part of our current understanding of nature and are at the heart of many phenomena.  However, the classical analog of fermionic mechanics is not as well known, because all the observables encountered in classical physics are bosonic. The purpose of this paper is to present the theory of the classical mechanics of fermions, sometimes called pseudo-classical mechanics.  Its usefulness is not in describing new physics, but in clarifying the mathematical structure and thus improving our understanding of the quantum theory. This theory was originally studied and formulated by Berezin\cite{Berezin}\cite{Berezin+Marinov}, Casalbuoni\cite{Casalbuoni}, and others, although so far its presentation has relied heavily on naive treatments of linear algebra and unwarranted analogies with the bosonic case, such that the true mathematical structure is obscured.
\section{Theory}
\begin{definition}
A \emph{fermionic phase space} is an orthogonal vector space, namely a vector space $\Phi$ with a positive-definite symmetric bilinear form $g$, also known as an inner product.
\end{definition}
This paper will consider fermionic phase spaces over $\mathbb{R}$. The generalization to complex fermions proceeds similarly to the generalization to complex bosons.  
\begin{definition}
A \emph{canonical transformation} is a diffeomorphism $U:\Phi\rightarrow\Phi$ that preserves the inner product in the sense of $U_*g=g$, which means $\forall\eta\in\Phi$, the derivative of $U$ is an orthogonal linear map:
\begin{equation}\label{eqn:canonical}
{D(U)^a}_b {D(U)^c}_d g_{ac} = g_{bd} .
\end{equation}
Dynamical evolution is assumed to be a one-parameter family of such transformations. Here and throughout I use abstract index notation for tensors over $\Phi$.
\end{definition}
\begin{theorem}\label{thm:linear}
Every canonical transformation is affine\footnote{There is a subtlety due to cases like the nonlinear Schr\"{o}dinger equation, which naively appear to be a counterexample to the Mazur-Ulam theorem.  The resolution probably lies in the presence of singular solutions, thus destroying the surjectivity of the evolution.  Either way, even though the nonlinear Schroedinger equation is norm-preserving, it still fails to preserve the inner product of an arbitrary pair of functions, and so cannot count as a canonical transformation.}.
\begin{proof}
The first step is to show that every canonical transformation is an isometry. For $\eta,\eta'\in\Phi$, the distance between them is equal to
\begin{equation}\label{eqn:metric}
d(\eta,\eta')=\underset{\gamma}{\min} \int_{0}^{1} \sqrt{\dot{\gamma}^a g_{ab} \dot{\gamma}^b} ,
\end{equation}
where the minimum is taken over all paths $\gamma:\mathbb{R}\rightarrow\Phi$ such that $\gamma(0)=\eta,\ \gamma(1)=\eta'$. $\dot{\gamma}$ denotes the tangent vector. When we transform with $U$,
\begin{equation}
d(U(\eta),U(\eta')) = \underset{\gamma}{\min} \int_{0}^{1} \sqrt{\dot{(U\circ\gamma)}^a g_{ab} \dot{(U\circ\gamma)}^b} ,
\end{equation}
where the minimum is still over the same set of paths as in \eqref{eqn:metric}.  The tangent vector of the transformed path $U\circ\gamma$ is
\begin{equation}
\dot{(U\circ\gamma)}^a = {D(U)^a}_b \dot{\gamma}^b ,
\end{equation}
so we have 
\begin{equation}
\dot{(U\circ\gamma)}^a g_{ab} \dot{(U\circ\gamma)}^b = {D(U)^a}_c \dot{\gamma}^c g_{ab} {D(U)^b}_d \dot{\gamma}^d = \dot{\gamma}^c g_{cd} \dot{\gamma}^d
\end{equation}
by the canonical property of U, equation \eqref{eqn:canonical}. So 
\begin{equation}
d(U(\eta),U(\eta'))=d(\eta,\eta') ,
\end{equation}
and thus $U$ is an isometry.  Since it is also surjective, it follows from the Mazur-Ulam theorem\cite{MazurUlam} that $U$ is affine.
\end{proof}
\end{theorem}
For fermions, we are usually only interested in canonical transformations that preserve 0, so in the remainder I will neglect translations and only consider linear canonical transformations. This restriction to linearity is in stark contrast to symplectic geometry, in which there are many nonlinear canonical transformations.\\
In particular, we have that $\forall\eta,\eta'\in \Phi$,
\begin{equation}
g(U(\eta),U(\eta') = g(\eta,\eta') ,
\end{equation}
so in fact $U$ is itself orthogonal.
\begin{theorem}
The generators of canonical transformations are anti-hermitian linear maps. Consequently, dynamical evolution must take the form
\begin{equation}
\frac{d\eta}{dt}=H(t)\eta ,
\end{equation}
where the Hamiltonian $H$ is a linear map $\Phi\rightarrow\Phi$, possibly depending on $t$, and $H^\dagger=-H$, where the adjoint is taken with respect to the inner product $g$.
\end{theorem}
The next step is to define the fermionic analog of the algebra of observables.  It is in fact a structure already well understood in mathematics: the exterior algebra\cite{exterioralg} of $\Phi$, sometimes also called the Grassmann algebra of $\Phi$.
\begin{definition}
A \emph{p-form} over $\Phi$ is a totally antisymmetric tensor of rank $p$ over $\Phi$. Let $\Lambda_p(\Phi)$ denote the space of $p$-forms over $\Phi$.  Given a $p$-form $\omega$ and a $q$-form $\alpha$, define their \emph{wedge product} to be the $p+q$-form
\begin{equation}
\omega\wedge\alpha\equiv \frac{(p+q)!}{p!q!} \omega^{[a...}\alpha^{b...]} .
\end{equation}
The \emph{algebra of observables} is the exterior algebra $\Lambda(\Phi)\equiv\bigoplus_{p=0}^n \Lambda_p(\Phi)$. The wedge product extends naturally to $\Lambda(\Phi)$ by bilinearity.
\end{definition}
\begin{theorem}
$\Lambda(\Phi)$ is an associative algebra under $\wedge$. It has a unit given by $1\in\Lambda_0(\Phi)$.
\end{theorem}
\begin{theorem}
$\Lambda(\Phi)$ is the unique associative algebra generated by $\Phi$ subject to the relations $\forall \eta,\eta'\in\Phi$,
\begin{equation}
\eta\wedge\eta' + \eta'\wedge\eta =0 .
\end{equation}
\end{theorem}
Given a $p$-form $\omega$ and a $q$-form $\alpha$,
\begin{equation}
\omega\wedge\alpha = (-1)^{pq} \alpha\wedge\omega .
\end{equation}
\begin{definition}
The \emph{Casalbuoni bracket} of a $p$-form $\omega$ and $q$-form $\alpha$ is the $(p+q-2)$-form
\begin{equation}
\{\omega,\alpha\} \equiv 2pq\ g_{ab}\omega^{[c...|a}\alpha^{b|d...]} .
\end{equation}
It extends naturally to a bilinear bracket on $\Lambda(\Phi)$.
\end{definition}
This structure gives the fermionic analog of the Poisson bracket on the algebra of observables. Given a $p$-form $\omega$, a $q$-form $\alpha$, and an $r$-form $\beta$, the Casalbuoni bracket satisfies
\begin{equation}
\{1,\alpha \}=0 ,
\end{equation}
\begin{equation}
\{\alpha,\omega \} = (-1)^{pq+1} \{\omega,\alpha \} ,
\end{equation}
\begin{equation}
\{\omega,\alpha\wedge\beta \} = \{\omega,\alpha \}\wedge\beta + (-1)^{pq} \alpha\wedge\{\omega,\beta \} ,
\end{equation}
\begin{equation}
(-1)^{pr}\{\omega,\{\alpha,\beta\} \} + (-1)^{qp}\{\alpha,\{\beta,\omega\} \} + (-1)^{rq}\{\beta,\{\omega,\alpha\} \} = 0 .
\end{equation}
A canonical transformation $U$ has a natural action as an automorphism of the algebra of observables, which on $p$-forms is given by
\begin{equation}
U(\omega)={U^{a_1}}_{b_1}\cdots{U^{a_p}}_{b_p} \omega^{b_1...b_p} .
\end{equation}
Note also that there is a canonical isomorphism between anti-hermitian linear maps and 2-forms:
\begin{equation}
H^{ab}={H^a}_c g^{cb} .
\end{equation}
Using this we can think of any generator $H$ of canonical transformations as a 2-form, and conversely.  The result is that we can express the action of the dynamical evolution on the observables using the Casalbuoni bracket: for $A\in\Lambda(\Phi)$,
\begin{equation}
\frac{dA}{dt}=\frac{1}{4} \{H,A \} .
\end{equation}
In particular, an arbitrary generator $G$ will satisfy
\begin{equation}
\frac{dG}{dt}=[H,G] ,
\end{equation}
where this is the commutator of linear maps.  This expresses Noether's connection between symmetry and conservation, in that the observable $G$ will be conserved if and only if the group generated by $G$ is a symmetry of $H$.\\
There are some additional structures that naturally arise on $\Lambda(\Phi)$. First, the inner product $g$ extends to an inner product on $\Lambda(\Phi)$:
\begin{equation}
\langle A,B\rangle\equiv \sum_{p=0}^{n} \frac{1}{p!} g_{a_1b_1}\cdots g_{a_pb_p} A_p^{a_1...a_p} B_p^{b_1...b_p} .
\end{equation}
If $\dim\Phi=n$, there is a natural $n$-form $\epsilon$, determined uniquely up to sign by the condition
\begin{equation}
\langle\epsilon,\epsilon \rangle=1 .
\end{equation}
The distinguished elements $1$ and $\epsilon$ have associated with them distinguished linear functionals:
\begin{definition}
The linear functionals $i,\int:\Lambda(\Phi)\rightarrow\mathbb{R}$ are
\begin{equation}
i(A)\equiv\langle 1,A\rangle ,
\end{equation}
\begin{equation}
\int A\equiv \langle\epsilon,A \rangle .
\end{equation}
\end{definition}
\begin{definition}
The \emph{derivative} is the linear map $\nabla:\Lambda(\Phi)\rightarrow\Lambda(\Phi)\otimes\Phi^*$ given by
\begin{equation}
\nabla A\equiv \{\cdot,A\} .
\end{equation}
On a $p$-form, this is
\begin{equation}
(\nabla_a\omega)^{b...} = g_{ac}\omega^{cb...} .
\end{equation}
\end{definition}
Note that the `derivative' defined here, and the `integral' defined above have nothing to do with differentiation or integration.  The notation is appropriate only because of the analogy with the bosonic case, in which the corresponding structures really are the derivative and integral.\\
This completes the mathematical structure of the classical theory of fermions.  The corresponding quantum theory alters this structure by ``deforming'' the wedge product on $\Lambda(\Phi)$ to a different associative product.  This deformed product is the Clifford product, determined uniquely by the relations $\forall \eta,\eta'\in\Phi$,
\begin{equation}
\eta\eta' + \eta'\eta =2g(\eta,\eta') .
\end{equation}
\begin{definition}
The \emph{quantum algebra of observables} is the Clifford algebra\cite{Clifford} of $\Phi$, which consists of $\Lambda(\Phi)$ with the Clifford product.
\end{definition}
\begin{theorem}
(Wick's theorem\cite{Wick}) The Clifford product of a set of vectors $\{\eta_i\in\Phi\}$ is given by 
\begin{equation}
\eta_1\cdots \eta_k = \sum_{\text{contractions}} (-1)^\sigma  \eta_1\wedge...\langle\text{contractions}\rangle...\wedge \eta_k .
\end{equation}
That is, it is a sum of wedge products of the $\eta_i$.  The sum is over all possible ways of contracting vectors, where a contraction of two vectors is just their inner product $g(\eta_i,\eta_j)$.  The sign $(-1)^\sigma$ is the sign of the minimal permutation needed to bring the contracted vectors next to each other.
\end{theorem}
The Clifford product is a deformation of the wedge product in the following sense: if we rescale the inner product $g$ by a parameter $\hbar$, then the Clifford product varies smoothly, and for $A,B\in\Lambda(\Phi)$,
\begin{equation}
AB|_{\hbar=0} = A\wedge B
\end{equation}
and
\begin{equation}
\frac{d}{d\hbar}AB|_{\hbar=0} =\frac{1}{2} \{A,B\} .
\end{equation}
The Clifford product has an associated involution $\dagger$ defined by 
\begin{equation}
\omega^\dagger\equiv (-1)^{\frac{1}{2}p(p-1)} \omega
\end{equation}
for $p$-forms. It satisfies
\begin{equation}
(AB)^\dagger = B^\dagger A^\dagger ,
\end{equation}
\begin{equation}
\langle A,B \rangle = i(A^\dagger B) ,
\end{equation}
\begin{equation}
\langle A,BC\rangle = \langle B^\dagger A,C\rangle ,
\end{equation}
\begin{equation}
\langle A,B\rangle=\langle A^\dagger, B^\dagger\rangle .
\end{equation}
Finally, the quantum dynamical evolution is an automorphism of the Clifford algebra.  For a theory with classical Hamiltonian $H$, the corresponding quantum evolution is given by
\begin{equation}\label{eqn:evolution}
\frac{dA}{dt}=\frac{1}{4}[H,A] ,
\end{equation}
where now the commutator is the Clifford algebra commutator.\\
In the classical fermionic case, all canonical dynamical evolutions are linear, with the Hamiltonian being a 2-form. The transition to quantum dynamics opens the possibility of allowing Hamiltonians of higher degree, which can be interpreted as including interactions.  We think of the evolution as an automorphism of the Clifford algebra, still obeying equation \eqref{eqn:evolution}, but letting the Hamiltonian be an arbitrary anti-hermitian element of the algebra.  However, such a theory would abandon the preservation of the phase space structure, and with it the possibility of interpreting the theory in terms of a corresponding classical evolution.

\end{document}